# THE THEORETICAL DFT STUDY OF ELECTRONIC STRUCTURE OF THIN Si/SiO$_2$ QUANTUM NANODOTS AND NANOWIRES

Pavel V. Avramov[1, 2], Alexander A. Kuzubov[2], Alexander S. Fedorov,[2] Pavel B. Sorokin[2, 3], Felix N. Tomilin[2] and Yoshihito Maeda[1, 4]

[1] Takasaki-branch, Advanced Science Research Center, Japan Atomic Energy Agency, Takasaki, 370-1292, Japan

[2] L.V. Kirensky Institute of Physics SB RAS, Academgorodok, 660036, Russia

[3] N.M. Emanuel Institute of Biochemical Physics of RAS, 119334 Moscow, Russian Federation

[4] Department of Energy Science and Technology, Kyoto University, Sakyo-ku, Kyoto 606-8501, Japan

The atomic and electronic structure of a set of proposed thin (1.6 *nm* in diameter) silicon/silica quantum nanodots and nanowires with narrow interface, as well as parent metastable silicon structures (1.2 *nm* in diameter), was studied in cluster and PBC approaches using B3LYP/6-31G* and PW PP LDA approximations. The total density of states (TDOS) of the smallest quasispherical silicon quantum dot ($Si_{85}$) corresponds well to the TDOS of the bulk silicon. The elongated silicon nanodots and 1D nanowires demonstrate the metallic nature of the electronic structure. The surface oxidized layer opens the bandgap in the TDOS of the Si/SiO$_2$ species. The top of the valence band and the bottom of conductivity band of the particles are formed by the silicon core derived states. The energy width of the bandgap is determined by the length of the Si/SiO$_2$ clusters and demonstrates inverse dependence upon the size of the nanostructures. The theoretical data describes the size confinement effect in photoluminescence spectra of the silica embedded nanocrystalline silicon with high accuracy.

[*]Correspondent author,

e.mail: avramov.pavel@ jaea.go.jp

+81 27 346 9670

+81 27 346 9696

# I. INTRODUCTION

Owing to their potential applications in optoelectronics, the study of silicon nanowires and quantum dots is a very active field of research. Optical properties of these confined systems are known to be quite different from their bulk counterparts and even from each other. The specific role of the surface oxidized layer [1,-7] (or Si/SiO$_2$ interface) in formation of the optical properties of the species (see, for example, work of Ref. 8) has not been studied theoretically yet because of the complexity of the atomic structure of the silica embedded nanocrystalline silicon particles.

Zero- or one-dimensional silicon and silicon-silica quantum dots (QD) and nanowires (NW) of various shapes and structures were synthesized under high temperature conditions (see, for example, works of Ref. 1-7, 9-18). The bulk-quantity Si nanowires covered by an SiO$_2$ layer [1,2] were synthesized by thermal evaporation of powder mixtures of silicon and SiO$_2$ or by thermal evaporation of silicon monoxide. [9] The polycrystalline nature of the silicon core was clearly seen [1-2, 9] on the TEM images of the species. The laser ablation in combination with the cluster formation and vapor-liquid-solid growth (VLS) method [11] was used to synthesize the bulk quantities of fine (6-20 *nm* in diameter) Si/SiO$_2$ nanowires (1 - 30 micrometers in length) using metallic iron as a catalyst. The TEM images demonstrated the perfect crystalline structure of the silicon core of the Si/SiO$_2$ objects, as well as the amorphous nature of the outer silica layer with perfectly round cross-section.

The short survey paper [13] demonstrates a number of perfect silicon NW structures with different sizes and shapes from small round and polygonal cross-sections up to square ones for structures with relatively big characteristic thickness. This fact can be explained by a strong surface tension of the thin species that changes the crystalline nature of the SiNWs. For the larger sizes, the surface tension becomes less important and can not change the cross-section of the NWs, which keeps the bulk nature of the atomic structure.

The structure and properties of silicon quantum dots with characteristic diameters from 2 to 10 *nm* embedded into silica environment was studied, for example, in the works of Ref. 3-7. In contrast with the semiconducting nature of the bulk silicon, the metallic behavior of the columns formed by nan1ocrystalline silicon (*nc*-Si) particles was detected.[3] Oxidation of the columns leads to the formation of *nc*-Si (with the characteristic size ~2*nm*) superlattice embedded into amorphous silica environment. The metallic conductivity was attributed to the significant structural tension in the interface regions. The decreasing of the tension by hydrogen chemosorption or by the creation of the oxidized layer on the surface of the species recovers the semiconducting nature of the silicon nanoobjects.

The structure of ultrathin $Si(111)/SiO_2$ interfaces was studied in the work of Ref. 19 by photoelectron spectroscopy. It was shown that before annealing, the interface thickness is ~ 2 monolayers of silicon whereas the annealing makes it more abrupt (~1.5 monolayers). The Si2*p* XPS spectrum [6] of the *nc*-Si quantum dots embedded into the $\alpha$-$SiO_2$ matrix (the thickness of the surrounding $\alpha$-$SiO_2$ layer is estimated ~1.6 *nm* [6]) demonstrates only $Si^{+4}$ ($SiO_2$) and $Si^0$ (*nc*-Si) components. It is the direct evidence of the real sharpness of the interface.

The photoluminescence (PL) properties of the nanosized silicon were widely studied in the literature (see, for example, [6-8, 20-29] including a detailed review [20]). The strong size confinement dependence of the PL photon energy upon the effective size of the *nc*-Si parts (or, exactly speaking, upon the confinement parameter $1/d$, where $d$ is the characteristic size of the *nc*-Si parts) was demonstrated for nanocrystalline silicon from 1 [21] and up to 9 [7] *nm*. For the different samples, the PL photon energy can vary from 2.30 eV (with 1 *nm* effective size [21]) and up to 1.28 eV (4.5 *nm* [23]). It was shown that the role of surface effects in the determination of the PL properties of the *nc*-Si structures is really important. [8, 22, 27, 30]

The luminescence properties of the silicon nanowires have been studied experimentally in significantly less extent. [28, 29] It was shown, [28] through the production of silicon nanowires by the thermal vaporization of the Si(111) substrate, that the quantum confinement effect increases the width of bandgap from 1.1 eV (bulk silicon) up to 1.56 eV. The electroluminescence peak with the energy 600 *nm* (2.07 eV) [29] was occurred from silicon nanowires with average diameter 4 *nm* as a result of band-to-band electron-hole recombination.

Some possible pristine 0D and 1D silicon nanostructures without saturation of the silicon surface dangling bonds were studied theoretically [31-38] using a set of *ab initio* electronic structure methods and molecular dynamics simulations. The atomic structure of the most promising small 0D and 1D polygonal nanoobjects was proposed in [31] and [34] based on the tetrahedral or 1D trigonal prism fragments of the bulk silicon selected along the [110] direction. In the calculations, a rearrangement of the surface of the objects due to dimer formation was found. The DFT electronic structure calculations of the 1D silicon structures reveal the metallic or semimetallic nature of the electronic structure of the objects. [35, 37]

The atomic and electronic structure of 0D and 1D silicon structures covered by hydrogen were studied in the works of Refs. 31, 38-48 using TB, [39, 42, 43] empirical Hamiltonian direct diagonalization [38] and LDA [31, 40, 41, 44-48] approaches. To calculate optical properties the Effective Mass Approach (EMA), [38, 42] dipole approximation, [41, 43] Green Wave function (GW) [46, 48] and Time Dependent DFT (TD DFT) [45, 47] approximations were used. The atomic models of the 0D and 1D species were based on the crystal structure of bulk silicon with square or rectangular crossections. All electronic structure calculations demonstrate a semiconducting forbidden gap and describe well the quantum confinement effect.

The study of the role of surface oxygen in the electronic structure of the *nc*-Si was performed in significantly less extent (see [8, 49-52]). Only in the work of Ref. 8 the electronic structure of some small silicon clusters covered by hydrogen with one Si=O bond on the surface was calculated using cluster LDA approach. Based on the LDA results, some large silicon structures (~5 *nm* effective size) covered by hydrogen with one Si=O bond on the surface were calculated using TB approach. Even one single surface Si=O bond leads to a considerable red shift of the bandgap in the electronic structure of the objects.

As it has been shown in the Introduction section, no realistic atomic models of the Si/SiO$_2$ quantum dots and nanowires with narrow Si/SiO$_2$ interfaces - taking into account the silica environment of the *nc*-Si core - have been developed and studied by *ab initio* technique. This is in spite of the electronic structure of pristine silicon and hydrogen passivated silicon nanostructures studied in detail through a set of experimental and theoretical approaches. A molecular design of the Si/SiO$_2$ quantum dots and nanowires combined with *ab initio* DFT and LDA PBC calculations of the species, is the main goal of the present paper.

## II. OBJECTS UNDER STUDY AND METHODS OF THE ELECTRONIC STRUCTURE CALCULATIONS

According to the work of Ref. 34, the pentagonal SiNWs are the most stable nanowires among all 1D structures with small diameter. This important theoretical result can be indirectly confirmed by STM images (work of Ref. 13 and references therein) of thin SiNWs which have almost polygonal or spherical shapes (in contrast [13] with the middle-size or thick particles which reveal the square or rectangular cross-sections). Due to these facts we used the thin pentagonal silicon nanowires [34] as the background to develop the atomic models of silicon and silicon/silica 1D nanostructures. To study the electronic structure of the objects and to understand the role of oxidized surface layer, we performed the electronic structure calculations of the species in cluster and PBC approximations.

Like in the work of Ref. 34, the silicon nanowires and nanoparticles with pentagon cross-section were designed by cutting out the prisms along [110] direction from the bulk silicon with one (100) and two {111} planes forming a nearly $2\pi/5$ angle. Therefore, with little shear strain and inexpensive {111} stacking faults, five such prisms can form a pentagonal wire, where all (100) facets permit low-energy reconstruction with characteristic dimer-row pattern. All designed 1D silicon clusters (Fig. 1) have the same diameter (1.2 *nm*). All nanoclusters in Fig. 1 are presented in two projections. In comparison, a fragment of ideal crystalline structure of silicon (Fig. 1a), calculated using periodic boundary conditions is presented. The silicon nanoclusters contain 85 (Fig. 1b), 145 (Figs. 1c and 1d) and 265 (Figs. 1e and 1f) atoms. To decrease the surface tension of the last structures (Figs. 1c-1f) a cap-shaped structure of the tips was formed by cutting the silicon prisms along the [110] directions from both sides with formation of correspondent {111} surfaces. The length of the elongated clusters are equal to 1.3, 1.8 and 3.7 $nm$. The 145 and 265 atom structures can form two types of surface dimers perpendicular to the main axis of the nanowires with the decreasing (Figs. 1c and 1e) and increasing (Figs. 1d and 1f) of the surface. The PBC calculations were performed using two slab models (Figs. 1g, the low surface isomer and 1h, the high surface isomer) with 30 atoms each developed based on the results of the cluster calculations.

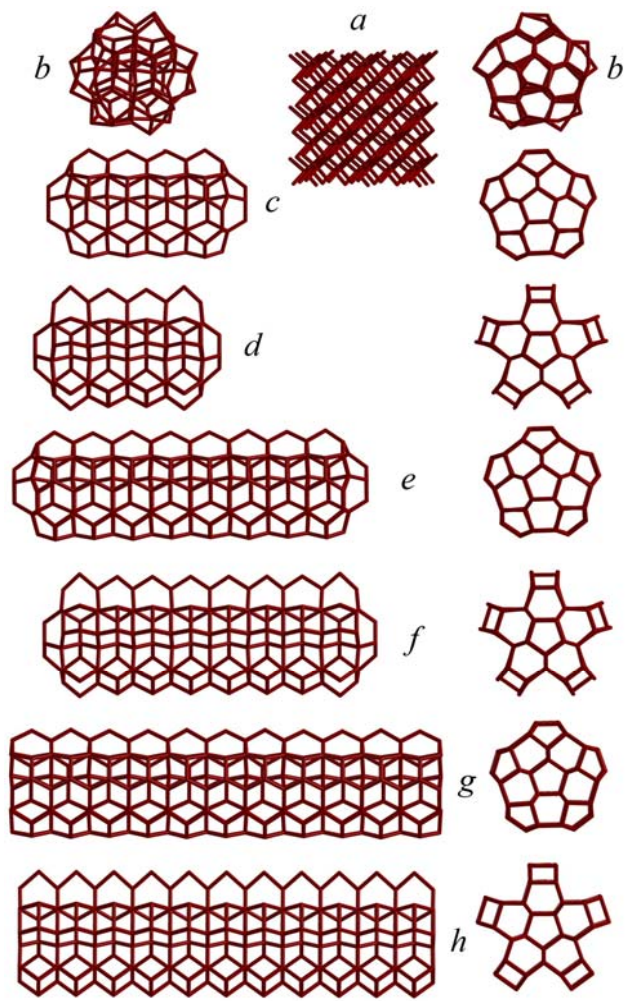

Fig. 1. (Color online) Atomic structure of the pristine silicon 0 and 1D nanoclusters. a) The fragment of the silicon crystal. b) Two projections of the $Si_{85}$ quantum dot. c) Two projections of the low energy $Si_{145}$ elongated nanocluster. The surface dimers are formed perpendicular to the main axis of the nanocluster to reduce the surface of the object. d) Two projections of the high energy $Si_{145}$ elongated nanocluster. The surface dimers are formed perpendicular to the main axis of the nanocluster to increase the surface of the object. e) and f) Two projections of the low and high energy $Si_{265}$ elongated nanoclusters with decreasing and increasing of the surface of the object. g) and h) Two projections of the low and high energy silicon infinite nanowires with decreasing and increasing of the surface of the object.

The atomic structure of the $Si/SiO_2$ nanoclusters (Fig. 2, all $Si/SiO_2$ clusters are presented in two projections. In comparison, a fragment of crystal lattice of α-quartz is presented in Fig 2a) was designed by connecting the silicon atoms with dangling bonds (these bonds form silicon dimers on the surface of pristine silicon nanowires, see above) on the (100) surface of the silicon core by bridged oxygen atoms. The {111}/(100) edges of different silicon prisms were connected to each other by the $SiO_2$ fragments with bridged silicon atoms between the $SiO_2$ fragments. For the thin $Si/SiO_2$ nanowires, the structural tension of the $Si/SiO_2$ interface is not big because of the relatively good matching of the structural parameters of silicon surface and Si-O chemical bonds. Such model of the $Si/SiO_2$ interface allows all surface related silicon atoms to keep the coordination number 4 which is natural to the $SiO_2$ oxide. The $Si/SiO_2$ interface of the caps of finite clusters was designed based on the same model. Due to the shape of the silicon core each cap has 5 silicon atoms with 3 Si-O bridged bonds. We guess that this fact is not critical for our study. It is well known (see, for example, [53] and references therein) that silicon-oxygen clusters often have nonstochiometric composition.

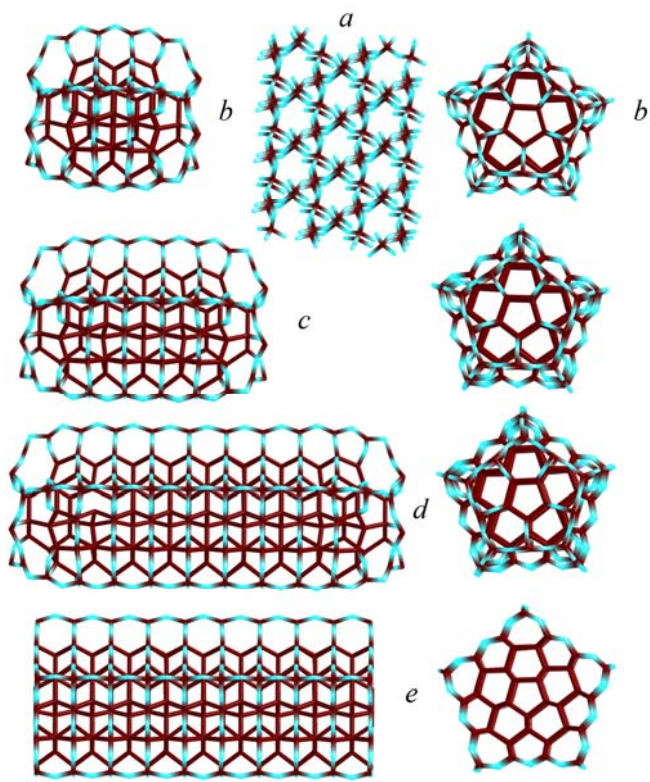

Fig. 2. (Color online) Atomic structure of the silicon/silica 0 and 1D nanoclusters. a) The fragment of the bulk α-quartz crystal. b) Two projections of the $Si_{125}O_{105}$ quantum dot. c) Two projections of the $Si_{195}O_{145}$ elongated cluster. d) Two projections of the $Si_{335}O_{225}$ elongated cluster. e) Two projections of silicon/silica infinite nanowire (the $Si_{36}O_{20}$ unit cell). The silicon atoms are presented in black (brown online), the oxygen atoms are presented in gray (light blue online).

The coverage of the smallest silicon cluster (85 silicon atoms) by oxidized layer leads to $Si/SiO_2$ structure with 230 atoms ($Si_{125}O_{105}$, Fig. 2b). The elongated $Si/SiO_2$ clusters contain 340 (on the base of initial $Si_{145}$ cluster, $Si_{195}O_{145}$, Fig. 2c) and 560 (on the base of initial $Si_{265}$ cluster, $Si_{335}O_{225}$, Fig. 2d) atoms. For the PBC calculations, the slab model with 56 atoms ($Si_{36}O_{20}$, Fig. 2e) was chosen.

To calculate the electronic structure of silicon and $Si/SiO_2$ nanoparticles in cluster approximation, the *GAMESS* [54] code was used. With one exception, the electronic structure calculations of all clusters including geometry optimization was performed at B3LYP/6-31G(d) level of theory [55] (from 1615 and up to 5880 basis set functions or 4420 - 14200 primitive Gaussians for the set of cluster described above) and using the analytic energy gradients. For the $Si_{335}O_{225}$ cluster, the atomic structure optimization was performed at B3LYP/3-21G(d) level (8390 basis functions or 14430 primitive Gaussians) and finally the DOS of the specie was calculated using 6-31G(d) basis set (9740 basis functions or 23720 primitive Gaussians). The closed shell electronic structure calculations of all structures except the $Si_{85}$ cluster were performed using the $D_{5h}$ point group of symmetry. For the smallest $Si_{85}$ structure, no symmetry restrictions were used because of possible significant distortion of the system by the surface tension. Plots of the calculated DOS were obtained using a broadening energy parameter of 0.1 eV. The resulting bandgaps have been extracted as the energy difference between two points of the DOS with intensities less than 0.1%.

The LDA approximation has previously been successfully used to describe the electronic structure of the silicon periodic systems (see, for example [56, 31, 35]). To compare the electronic structure of 1D periodic systems with literature data we used the LDA approximations to compute the electronic structure of the species within the PBC approximation. For all calculations of the periodic systems, the Vienna *ab-initio* simulation package (VASP) [57, 58] was used. The total energy code is based on LDA approximation [59, 60] with the Ceperley-Alder exchange-correlation functional, [61] plane-wave basis sets and ultrasoft pseudopotentials (PP). [62] The pseudopotentials allow one to significantly reduce the maximal kinetic energy cutoff without loss of accuracy. To calculate the electronic structure of all periodic systems we used 128 points of the *k*-space with $E_{cutoff}$=250 eV. All electronic structure calculations of the infinite silicon and silicon/silica nanowires were performed within the PBC approximation (LDA PW PP PBC) without symmetry restrictions of the unit cells. The geometry optimization was carried out until the forces acting on all atoms became lower 0.05 eV/Å. The same procedure to build up the DOSes and extract the bandgap values, as was described above for the cluster calculations, has been used.

### III. RESULTS AND DISCUSSION

*a) The pristine silicon structures*

The main factor affecting the atomic structure of the thin silicon nanowires is a surface tension caused by dangling bonds on the surface of the objects. The initial atomic structure of the smallest 0D $Si_{85}$ cluster was created using five triangular prisms by the procedure described in the previous Section. The surface tension (the B3LYP/6-31G* optimized atomic structure, Fig. 1b) leads to a formation of close to spherical $Si_{85}$ quantum dots, keeping the pentagonal nature of the cluster with mirror symmetry along the *xy* plane.

Taking into account a conservation of the symmetry of the smallest cluster, the atomic structure optimization of all other silicon clusters was performed using the $D_{5h}$ symmetry restrictions. As it was mentioned above, the surface relaxation of crystalline silicon occurs through formation of the surface dimers. For the 1D silicon structures there are two ways of dimer formation: parallel and perpendicular to the main axis of an object. In case of parallel formation of dimers the atomic structure of the clusters become unstable and undergo fragmentation on small spherical particles (85 + 60 atoms, for example) during atomic structure optimization. The main reason of the fragmentation is an involvement of tip silicon atoms into the dimer formation process.

The dimer formation perpendicular to the main axis of the elongated clusters with the decreasing of surface of silicon nanostructures (Figs. 1c and 1e) leads to the low energy isomers, whereas the increasing of surface (Figs. 1d, 1f) leads to the high energy isomers. On the B3LYP/6-31G* level, the first $Si_{145}$ isomer is 0.05 eV/atom more preferable than the second one. For the $Si_{265}$ pair, this difference is bigger (0.08 eV). In both cases the tips of the clusters are not involved in the dimer formation process. Because of this, the bigger the length of the clusters, the bigger the energy difference between isomers.

The optimization of the atomic structure of the infinite silicon nanowires made by LDA PW PP PBC calculations without symmetry restrictions confirms the results of cluster B3LYP/6-31G* calculations. The atomic structure of low (Fig. 1g) and high (Fig. 1h) energy isomers of the thin silicon nanowires was determined. The decreasing of the surface of the nanowires leads to the lowering of the energy of the system. The energy difference between the isomers is equal to 0.1 eV/atom. The significant increasing of this value in comparison with the $Si_{145}$ (0.05 eV/atom) and $Si_{265}$ (0.08 eV/atom) can be explained by the absence of the tips of the objects in the PBC calculations.

The DOS of the 0 and 1D structures calculated using the B3LYP/6-31G* and LDA PW PP PBC methods, as well as the DOS of the bulk silicon (LDA PW PP PBC calculation) are presented in Fig. 3. The DOS of the $Si_{85}$ quantum dot (Fig. 3b) is close to the shape of the DOS of bulk silicon (Fig. 3a), but with small finite density of electron states at the Fermi level. The main difference between the DOSes is a presence of two small intensity peaks from both sides of the Fermi level with -0.3 and 0.3 eV energies respectively. In general, the DOS of the $Si_{85}$ cluster reproduces well (relative peak energies and intensities) the DOS of the bulk silicon in wide energy region from -8 and up to 2 eV (Figs. 3a, 3b).

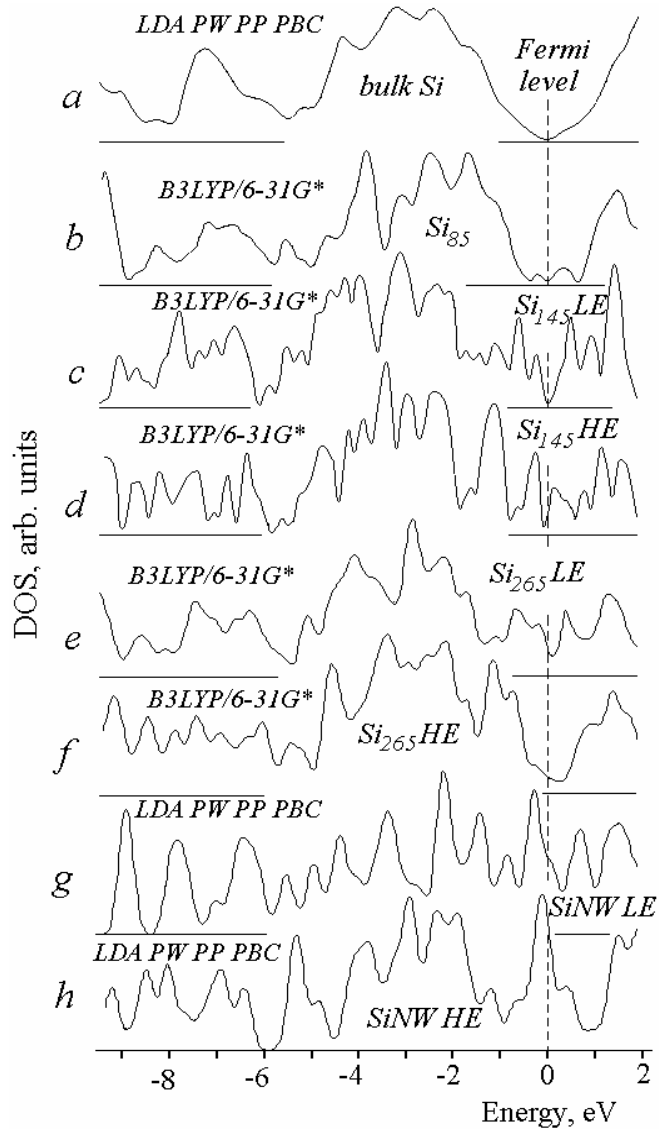

Fig. 3. The electron density of states of a) Bulk silicon LDA PW PP PBC DOS. b) $Si_{85}$ B3LYP/6-31G* DOS. c) The B3LYP/6-31G* DOS of low energy (LE) isomer of elongated $Si_{145}$ cluster. d) The B3LYP/6-31G* DOS of high energy (HE) isomer of elongated $Si_{145}$ cluster. e) The B3LYP/6-31G* DOS of low energy isomer of elongated $Si_{265}$ LE cluster. f) The B3LYP/6-31G* DOS of high energy isomer of elongated $Si_{265}$ HE cluster. g) The LDA PW PP PBC DOS of the low energy isomer of infinite pristine silicon nanowire. h) The LDA PW PP PBC DOS of the high energy isomer of infinite pristine silicon nanowire.

Elongation of the clusters (Fig. 3 c-h) leads to fundamental changes in the DOSes of the species. All objects have the metallic nature of the electronic structure. The bigger the length of the objects, the higher the DOS at the Fermi level. Only low energy isomers of $Si_{145}$ and $Si_{265}$ clusters have close energy peak positions in the density of electron states in the energy region -2.5 ÷ -8 eV. All other structures (high energy isomers of $Si_{145}$ and $Si_{265}$ clusters (Figs. 3d, f) and both isomers of the infinite silicon nanowires (Fig. 3g, h)) have completely different DOS shapes. The most fundamental changes of the DOSes can be observed at the Fermi level. The higher the surface tension of the species (because of the number and types of surface dimers), the higher the DOS at the Fermi level. This fact clearly demonstrates the leading role of the surface tension in the formation of metallic conductivity of the silicon species with unpassivated or strained surface.

The caps of the finite species are not involved in the process of dimer formation perpendicular to the main axis of the objects. The bigger the length of the nanoclusters, the smaller the relative weight of the terminal cap structures. The surface dimer structures lead to a significant distortion of the atomic structure of the silicon clusters (Fig. 1), especially in the case of high energy isomers. The surface dimer formation and consequent atomic structure distortion of the species are the main reasons of fundamental changes in the electronic structure of silicon 0 and 1D nanostructures.

*a) The silicon/silica structures*

The DFT optimization of the atomic structure of silicon/silica objects in both cluster and PBC approximations clearly demonstrates the stabilization role of the surface oxidized layer. The bridged oxygen atoms on the surface of silicon cores prevent a formation of the silicon dimer structures and considerable distortion of the atomic structure of the silicon cores of the $Si/SiO_2$ nanoclusters. The parameters of the Si-O chemical bonds in α-quartz (the SiO distance is equal to 1.298 Å and the O-Si-O angle is equal to 111.8°) can not exactly match the unrelaxed (100) surface of silicon. The cluster B3LYP/6-31G* and LDA PW PP PBC methods give practically the same parameters of Si-O bonds at the $Si/SiO_2$ interface. On the pentagon apexes (see both parallel and perpendicular projections of the infinite $Si/SiO_2$ nanowire, Fig. 2) the Si-O bond length is equal to 1.839 Å and the O-Si-O angle is equal to 149.1°. The lengths of two other Si-O bonds of the $SiO_2$ fragments are equal to 1.671 Å with the O-Si-O angle 126.3°. The bridged Si-O bonds that connect the $SiO_2$ fragments with the (100) silicon surface are equal to 1.684 Å with the O-Si-O angle 127.4°. Finally, the bond length of the bridged Si-O fragments on the (100) silicon surface is equal to 1.72 Å with the O-Si-O angle 127.5°. So, the Si-O bonds deviate from the bulk $SiO_2$ ones in the range 28-42% (bond lengths) and 13-33% (bond angles). Nevertheless, the structural strength of the $Si/SiO_2$ interface allows one the existence of the thin 0 and 1D silicon/silica nanoclusters. This can not be true for the large ones (see Introduction section) due to the large total surface tension caused by mismatching of the silicon and silica crystal lattices.

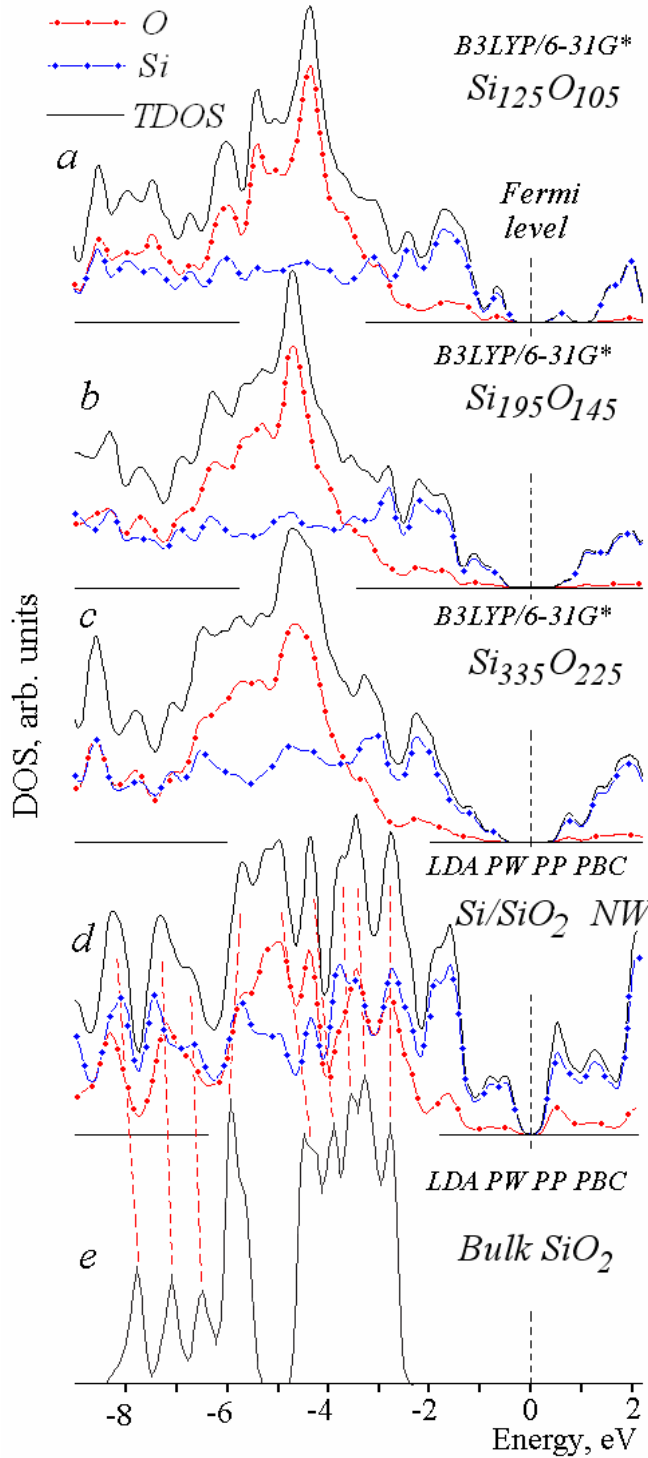

Fig. 4. (Color online) The electron density of states of a) $Si_{125}O_{105}$ quantum dot B3LYP/6-31G* DOS. b) The B3LYP/6-31G* DOS of the elongated $Si_{195}O_{145}$ cluster. c) The B3LYP/6-31G* DOS of the elongated $Si_{335}O_{225}$ cluster. d) The LDA PW PP PBC DOS of the infinite silicon/silica nanowire. e) The LDA PW PP PBC DOS of bulk α-quartz. The total density of states are presented by black solid lines. The partial oxygen DOSes are presented by (red online) dashed lines with filled (red online) circles. The partial silicon DOSes are presented by (blue online) dashed lines with filled (blue online) squares.

The total and partial silicon and oxygen DOSes of all 0 and 1D Si/SiO$_2$ species are presented in Fig. 4, as well as one of the α-quartz calculated using the B3LYP/6-31G* and LDA PW PP PBC methods. All 0 and 1D species have definitely pronounced semiconducting bandgaps (Fig. 4 a-d). A comparison of the experimental [7] and cluster B3LYP/6-31G* theoretical data (Table, Fig. 5) demonstrates an excellent agreement between the B3LYP/6-31G* theory and PL experiment. The theoretical data clearly confirms the size confinement effect [7, 8, 21, 23, 25] shortly described in the Introduction section. The bandgap of the finite Si/SiO$_2$ clusters varies from 2.2. eV for the 1.7 *nm* cluster (230 atoms) and up to 1.4 eV for the 3.8 *nm* cluster (560 atoms). The bandgap of the Si/SiO$_2$ infinite nanowire is equal to 0.6 eV (Fig. 4d), which is smaller than the bandgap of bulk silicon (1.1 eV) probably due to the well known underestimation of the semiconducting bandgaps by the LDA approximation. Comparison of the metallic nature of the pristine Si nanostructures (see previous paragraph and Fig. 3) with the electronic structure of thin Si/SiO$_2$ 0 and 1D nanoclusters clearly demonstrates the main role of the SiO$_2$ oxidized layer in the bandgap formation of the species. This effect can be explained by the neutralization of surface dangling bonds and the prevention of surface silicon dimer formation, low energy position of strongly localized Si-O chemical bonds and polarization of the electronic structure of the silicon core by negatively charged oxygen ions from the surface oxidized layer.

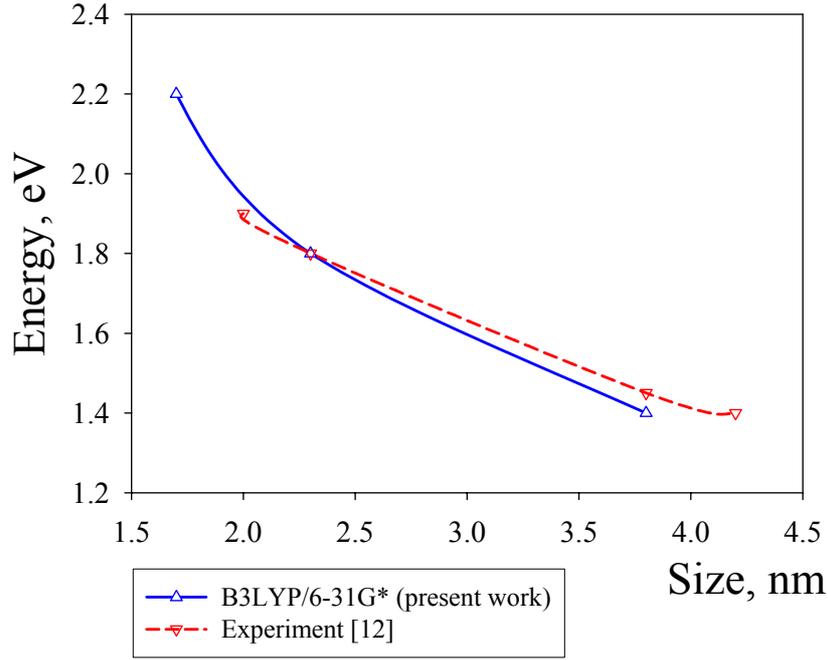

Fig. 5. The theoretical B3LYP/6-31G* and experimental [7] dependence of the bandgap of the *nc*-Si upon the confinement parameter 1/*d*, where *d* is the characteristic size of the clusters. The dark (blue online) line presents the theoretical B3LYP/6-31G* results whereas the dashed (red online) line presents the experimental data.

The excellent agreement between the experimental [7] and theoretical data can be explained by two reasons. Firstly, even for the smallest $Si_{125}O_{105}$ quantum dot, the total number of the basis functions (3950, or 9440 primitive Gaussians, 6-31G* basis set) is good enough for the correct description of the electronic structure of the object. For the large elongated clusters ($Si_{195}O_{145}$ and $Si_{335}O_{225}$), the correspondent numbers (5880/14200 and 9740/23720) are significantly larger. So, the large total number of the basis functions/primitive Gaussians in combination with the hybrid B3LYP DFT potential can be the main reason for the high accuracy of the DFT calculations of the $Si/SiO_2$ nanoclusters.

Table. Experimental PL [7] and Refs. therein and theoretical transition energies for the nc-Si/$SiO_2$ QDs

| Experiment [12] and Refs. in it | | | Theory (cluster B3LYP/6-31G*) | | |
|---|---|---|---|---|---|
| Size (*nm*) | PL1 (eV) | PL2 (eV) | Size (*nm*) | 1st transition (eV) | 2nd transition (eV) |
| 2.0 | -- | 1.8÷1.9 | 1.7x1.6 | 1.3 | 2.2 |
| 2.3 | -- | 1.8 | 2.3x1.6 | 1.3 | 1.8 |
| 3.8 | -- | 1.45 | 3.8x1.6 | 0.8 | 1.4 |
| 4.2 | 0.9 | 1.4 | | | |

Even the authors of the work of Ref. 7 claimed the spherical shape (or 0D nature) of the *nc*-Si parts; our DFT calculations of the 1D species describe well the experimental [7] size confinement effect. In the simplest approximation, the effect can be scaled using two parameters $1/r + 1/L$, where $r$ and $L$ are the radius and length of the clusters correspondingly. Since the size confinement effect reflects the inverse dependence of the bandgap upon the size of the species, the leading term in the expression is the $1/L$. The $1/r$ does not play any role in the determination of the bandgap since the electronic states derived by the $1/r$ term lie below or above the HOMO and LUMO levels derived by the leading $1/L$ term.

The DOS of infinite Si/SiO$_2$ nanowire (Fig. 4d) in the region of $-5 \div -8$ eV obtained by the LDA PW PP PBC method reflects the shape of $\alpha$-quartz LDA PW PP PBC DOS (Fig. 4e). Both DOSes have close relative peak intensities and energy positions. This fact can be explained by the localized character and high energy of the Si-O bonds. The top of the valence band (energy region $-5 - 0$ eV) of all Si/SiO$_2$ species mainly consists of the Si derived states of the silicon core (Fig. 4). The conductivity band of the infinite Si/SiO$_2$ nanowire (Fig. 4d) contains some small amount of oxygen density near the Fermi level. The analysis of the partial DOSes of the species shows that the Si derived states of the silicon core near the Fermi level (in both valence and conductivity bands) are responsible for the transport and optical properties of the species.

## IV. CONCLUSIONS

Based on the DFT calculations, we have shown that thin pristine silicon nanowires are metastable structures. The stability of the nanoclusters directly depends on the method of formation of (100) surface silicon dimers. The dimer formation parallel to the main axis of the nanowires leads to a fragmentation of the 1D nanowires and formation of the 0D silicon quantum dots. The metastable silicon 0D and 1D structures have finite density of electron states at the Fermi level. The surface tension and silicon dimers play the dominant role in the formation of metallic nature of electronic structure of the silicon nanoclusters. Because of the localized character of the Si-O chemical bonds, the oxidized layer opens an energy bandgap in the electronic structure of the Si/SiO$_2$ nanoclusters, preventing formation of the silicon dimers and decreasing of the surface tension. The DFT calculations clearly confirm the 1/$d$ size confinement effect in the optical spectra of the silica embedded Si-$nc$ species. The Si derived states of the silicon core determine the optical and transport properties of the silicon/silica objects.

## ACKNOWLEDGMENTS

This work was supported in parts by project "Materials Design with New Functions Employing Energetic Beams", JAEA, JAEA Research fellowship (PVA) and grants from the US Department of Energy via the Ames Laboratory and the Air Force Office of Scientific Research, Russian Fund of Basic Researches (grant n.05-02-17443), grant of Deutsche Forschungsgemeinschaft and Russian Academy of Sciences, No. 436 RUS 113/785 (PBS). The calculations have been partially performed on the Joint Supercomputer Center of the Russian Academy of Sciences. PVA also acknowledges Profs. Mark S. Gordon and K.-M. Ho for hospitality during stay his at Ames National Laboratory (USA) and the personnel of Research Group for Atomic-scale Control for Novel Materials under Extreme Conditions for hospitality and fruitful discussions. Authors also acknowledge Amir Fishman for technical help. The geometry of all presented structures was visualized by ChemCraft software [http://www.chemcraftprog.com].